\documentclass[iop]{emulateapj}

\def\ltsima{$\; \buildrel < \over \sim \;$}
\def\simlt{\lower.5ex\hbox{\ltsima}}
\def\gtsima{$\; \buildrel > \over \sim \;$}
\def\simgt{\lower.5ex\hbox{\gtsima}}
\def\kms{{\rm\,km\,s^{-1}}}

\def\kpc{{\rm\,kpc}}

\def\msun{{\rm\,M_\odot}}
\def\lsun{{\rm\,L_\odot}}

\def\AA{$\; \buildrel \circ \over {\rm A}$}

\def\deg{^\circ}

\def\s{\ifmmode \widetilde \else \~\fi}
\def\={\overline}

\def\spose#1{\hbox to 0pt{#1\hss}}

\def\lta{\mathrel{\spose{\lower 3pt\hbox{$\mathchar"218$}}
     \raise 2.0pt\hbox{$\mathchar"13C$}}}
\def\gta{\mathrel{\spose{\lower 3pt\hbox{$\mathchar"218$}}
     \raise 2.0pt\hbox{$\mathchar"13E$}}}
\def\Dt{\spose{\raise 1.5ex\hbox{\hskip3pt$\mathchar"201$}}}    
\def\dt{\spose{\raise 1.0ex\hbox{\hskip2pt$\mathchar"201$}}}    

\def\dotsfill{\leaders\hbox to 1em{\hss.\hss}\hfill}

\def\FeH{{\rm[Fe/H]}}

\shorttitle{Spectroscopy of Cas~III, Lac~I, \& Per~I}
\shortauthors{N. F. Martin et al.}

\begin{document}


\title{Spectroscopy of the three distant Andromedan satellites Cassiopeia~III, Lacerta~I, and Perseus~I}


\author{Nicolas F. Martin$^{1,2}$, Kenneth C. Chambers$^3$, Michelle L. M. Collins$^2$, Rodrigo A. Ibata$^1$, R. Michael Rich$^4$, Eric F. Bell$^5$, Edouard J. Bernard$^6$, Annette M. N. Ferguson$^6$, Heather Flewelling$^3$, Nicholas Kaiser$^3$, Eugene A. Magnier$^3$, John L. Tonry$^3$, Richard J. Wainscoat$^3$}

\email{nicolas.martin@astro.unistra.fr}

\altaffiltext{1}{Observatoire astronomique de Strasbourg, Universit\'e de Strasbourg, CNRS, UMR 7550, 11 rue de l'Universit\'e, F-67000 Strasbourg, France}
\altaffiltext{2}{Max-Planck-Institut f\"ur Astronomie, K\"onigstuhl 17, D-69117 Heidelberg, Germany}
\altaffiltext{3}{Institute for Astronomy, University of Hawaii at Manoa, Honolulu, HI 96822, USA}
\altaffiltext{4}{Department of Physics and Astronomy, University of California, Los Angeles, CA 90095-1547, USA}
\altaffiltext{5}{Department of Astronomy, University of Michigan, 500 Church St., Ann Arbor, MI 48109}
\altaffiltext{6}{Institute for Astronomy, University of Edinburgh, Royal Observatory, Blackford Hill, Edinburgh EH9 3HJ, UK}

\begin{abstract}
We present Keck~II/DEIMOS spectroscopy of the three distant dwarf galaxies of M31 Lacerta~I, Cassiopeia~III, and Perseus~I, recently discovered within the Pan-STARRS1 $3\pi$ imaging survey. The systemic velocities of the three systems ($v_\mathrm{r,helio} = -198.4\pm1.1\kms$, $-371.6\pm 0.7\kms$, and $-326\pm3\kms$, respectively) confirm that they are satellites of M31. In the case of Lacerta~I and Cassiopeia~III, the high quality of the data obtained for 126 and 212 member stars, respectively, yields reliable constraints on their global velocity dispersions ($\sigma_\mathrm{vr} = 10.3\pm 0.9\kms$ and $8.4\pm 0.6\kms$, respectively), leading to dynamical-mass estimates for both of $\sim4\times10^7\msun$ within their half-light radius. These translate to $V$-band mass-to-light ratios of $15^{+12}_{-9}$ and $8^{+9}_{-5}$ in solar units. We also use our spectroscopic data to determine the average metallicity of the 3 dwarf galaxies ($\FeH = -2.0\pm0.1$, $-1.7\pm0.1$, and $-2.0\pm0.2$, respectively). All these properties are typical of dwarf galaxy satellites of Andromeda with their luminosity and size.
\end{abstract}

\keywords{Local Group --- galaxies: individual: And~XXXI, And~XXXII, And~XXXIII, Cas~III, Lac~I, Per~I --- galaxies: kinematics and dynamics}

\section{Introduction}

The systematic coverage of the northern sky yielded by the Pan-STARRS1 (PS1) $3\pi$ survey provides a spatially complete mapping of the regions around Andromeda (M31), irrespective of the close proximity of the Galactic plane. This has led to the discovery of three dwarf galaxies: Cassiopeia~III (Cas~III or Andromeda~XXXII), Lacerta~I (Lac~I or Andromeda~XXXI; \citealt{martin13a}), and Perseus~I (Per~I or Andromeda~XXXIII; \citealt{martin13c}), all reasonably bright ($-12\simlt M_V \simlt -10$) and within $30\deg$ of M31. As they all three have M31-centric distances between 145 and $375\kpc$, they are thought to be distant satellites of M31. Such distant dwarf galaxies would have a significant weight in constraining the mass of their host \citep[e.g.][]{watkins10} or even its proper motion \citep{vandermarel08}. However, their usefulness in this endeavor depends on the confirmation that these three systems are indeed satellites of M31.

We have consequently embarked in a spectroscopic program to gather spectra of numerous stars in these systems with the DEIMOS spectrograph on Keck~II. Our goal is not only to determine the systemic velocity of the three dwarf galaxies to confirm that they are bound to M31, but also to infer their internal dynamics, masses, and metallicities, and how these compare to the properties of other (mainly closer) M31 dwarf galaxies.

In this Letter, we present the first results of our campaign and constrain the systemic velocities, velocity dispersions, and metallicities of the three dwarf galaxies. This work is structured as follows: Section~2 presents the observations and data reduction, Section~3 describes our analysis and results, and we discuss their implications in Section~4.

\section{Observations}
\begin{figure*}
\begin{center}
\includegraphics[width=0.32\hsize,angle=270]{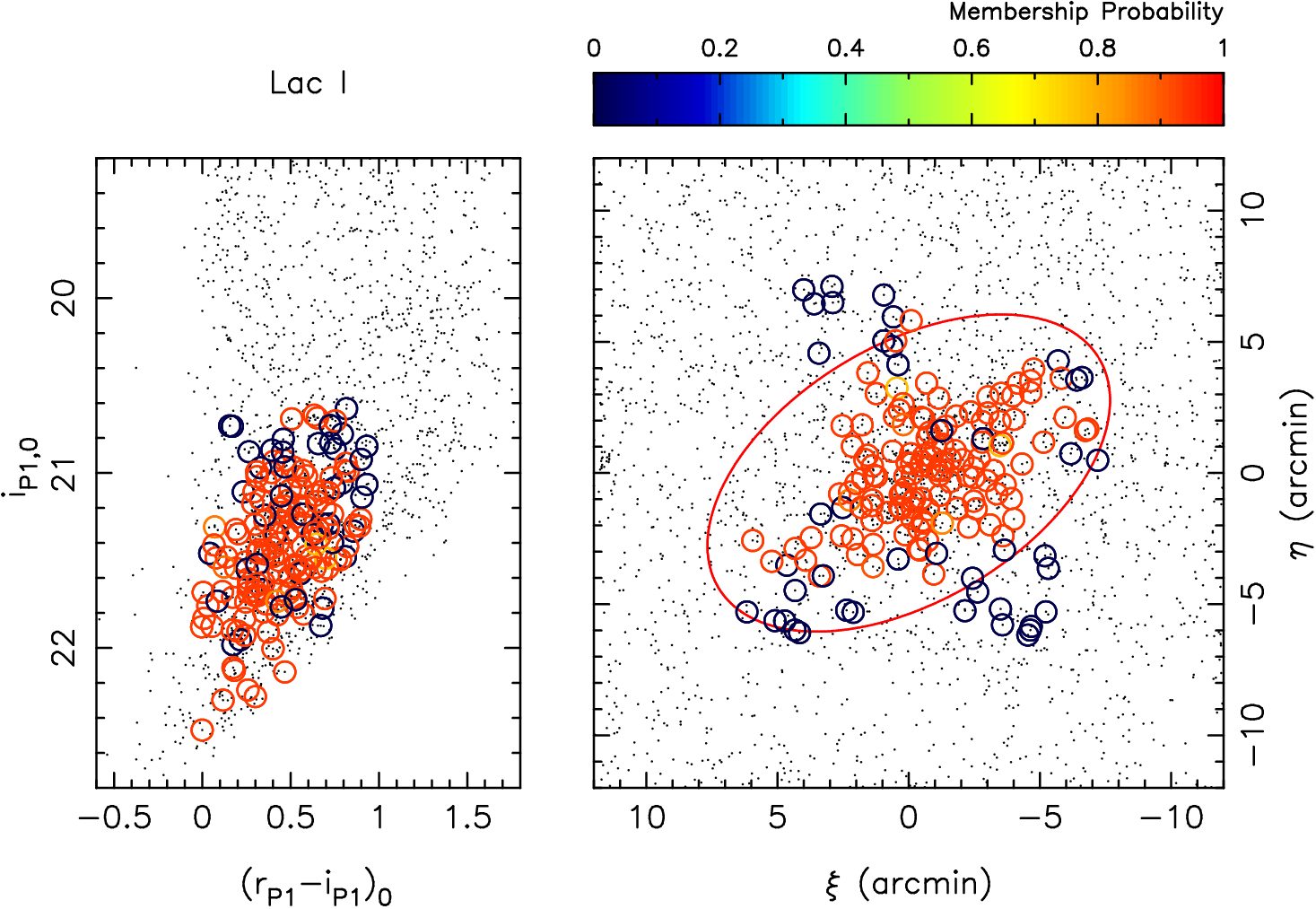}\hspace{1cm}
\includegraphics[width=0.32\hsize,angle=270]{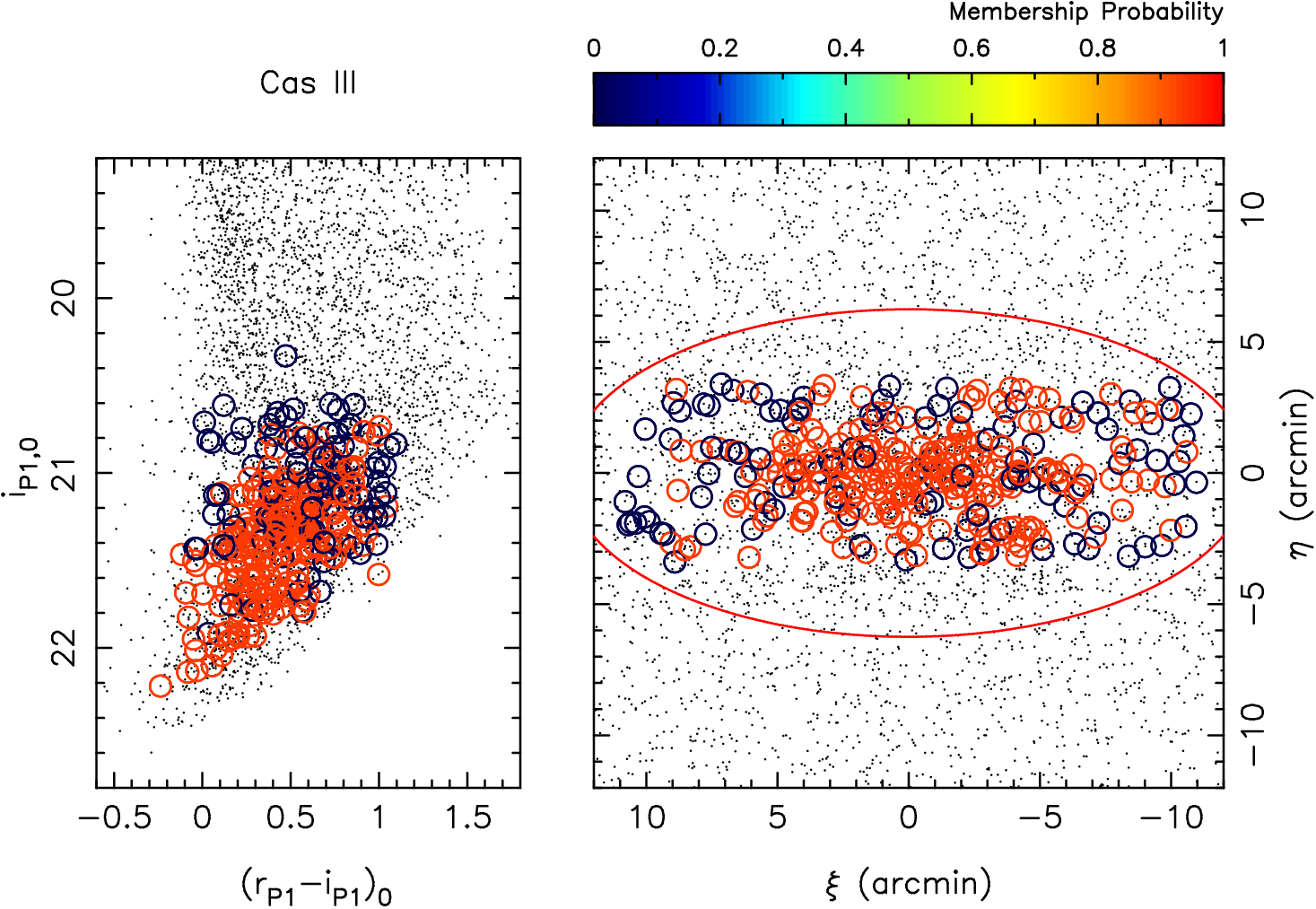}\vspace{0.5cm}
\includegraphics[width=0.32\hsize,angle=270]{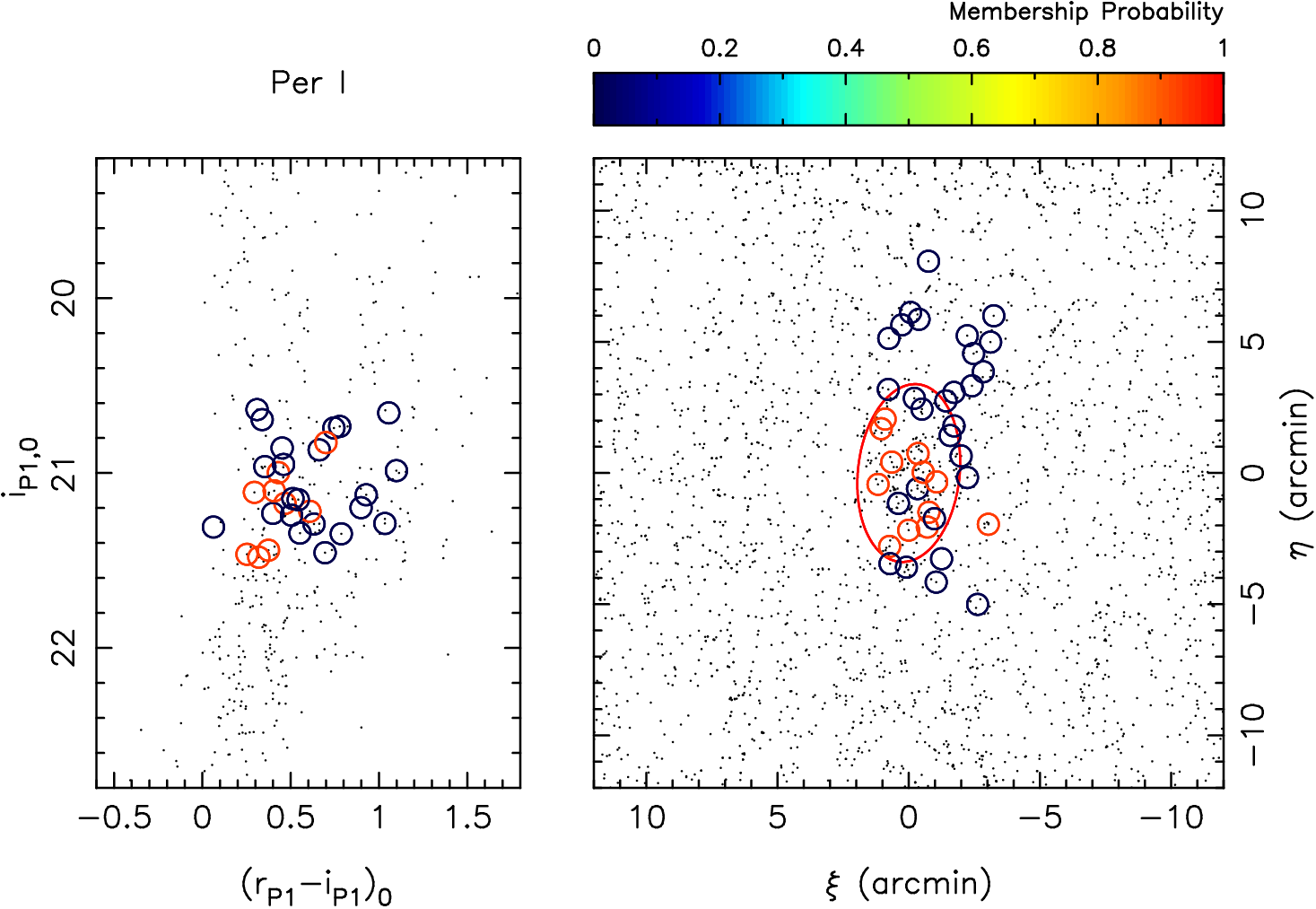}
\caption{\label{CMD_sp}Distribution of the targeted stars in the CMD and on the sky for Lac~I (top-left), Cas~III (top-right), and Per~I (bottom). The CMDs display all stars within $3r_h$ of the dwarf galaxies' centroids as black dots. The sky distributions show all stars loosely compatible with being RGB stars of the dwarf galaxies ($i_{\mathrm{P1},0}<20.5$ and $(r_\mathrm{P1}-i_\mathrm{P1})_0<1.1$) as black dots. The red ellipse indicates the region within $2r_h$ of the dwarf galaxies, following the structural parameters determined by \citet{martin13a} and \citet{martin13c}. Stars with good-quality spectra are circled with a color that corresponds to the membership probability as defined below in Section~\ref{sect:velocities}.}
\end{center}
\end{figure*}

The selection of targets for the DEIMOS observations is based on the PS1 photometry for Lac~I and Cas~III \citep{martin13a} and a rederivation of the photometry from the PS1 images in the case of Per~I \citep{martin13c}. Sources were selected to be compatible with red giant branch (RGB) stars from the three dwarf galaxies ($i_{\mathrm{P1},0}\simlt20.5$ and $(r_\mathrm{P1}-i_\mathrm{P1})_0\simlt1.1$) and priorities were set to favor stars closer to the tip of the RGB for which the photometric uncertainties are smaller. We created 2 masks for Lac~I, 4 for Cas~III, and 1 for Per~I. The resulting targets for which we acquired good quality spectra are highlighted in the color-magnitude diagrams (CMDs) and spatial distributions of Figure~\ref{CMD_sp}, color-coded by the dwarf-galaxy-membership probability we calculate in the next Section.

The Lac~I and Cas~III observations were conducted during the nights of August 28--29, 2013, under good to exquisite conditions (clear sky, seeing in the range $0.4$--$0.9''$). Each mask was observed for a total of 7200\,s, split into six 1200\,s subexposures. A realignment of the masks was performed after every subexposure to ensure that we could track the stability of a star's location inside its slit. Moreover, once on target, we observed NeArKrXe arc lamps before and after the acquisition of a mask to facilitate the wavelength calibration. The single Per~I mask was acquired on the night of October 3, 2013 since this galaxy was not yet discovered at the time of the other observations. This mask was observed for $3\times1200$\,s under reasonable seeing conditions ($\sim1''$), but with fleeting clouds. In addition, one of the two Lac~I masks was also reobserved during this night to check the stability of the setup and data reduction over time. For all the observations, the instrumental set up covers the 5600--9800\AA\ wavelength range with the 1200 line/mm$^{-1}$ grating at a spectral resolution of $\sim0.33$\AA/pixel. The Lac~I and Cas~III spectra have typical signal-to-noise per pixel $S/N>10$ at $i_\mathrm{P1}=21.0$ and $S/N>4$ at $i_\mathrm{P1}=22.0$. For the Per~I mask with a shorter integration time, $3<S/N<6$.

The spectra are extracted and reduced through a custom-built pipeline that follows the procedure described in \citet{ibata11a} and \citet{collins13}. In addition, the quality of our spectra is good enough that we can use the Fraunhofer A band line in the range 7595--7630\AA\ to calculate small telluric corrections that stem from minute misalignments of the stars in the slits \citep[e.g.][]{sohn07}. Heliocentric radial velocities, $v_\mathrm{r}$, are measured from the three strong lines of the Calcium triplet at 8498, 8542, and 8662\AA. We only keep in our sample stars for which $S/N>2$ per pixel, the uncertainty on the velocity, $\delta_{\mathrm{vr}}$, is smaller than $10\kms$, and $-600<v_r<200\kms$.

Our observations include a sizable sample of 82 good-quality repeat measurements, which allows for a reliable determination of the uncertainty floor of our observations, $\epsilon=3.4\pm0.5\kms$, as defined in equation (1) of \citet{simon07}. This uncertainty floor is added in quadrature to the measured velocity uncertainties, which also include the uncertainty from the telluric correction. Multiple observations of the same star are combined through the uncertainty-weighted average of their velocities before adding the uncertainty floor.

A visual inspection of the spectra betrays the presence of a handful of carbon stars in both Lac~I and Cas~III, identified from the characteristic ``serrated'' shape of their spectra. Although the detailed study of these stars is beyond the scope of this Laper, their presence could indicate that Lac~I and Cas~III host intermediate age populations. These carbon stars were removed from the data sets studied in this Letter.

\section{Results}
\subsection{Velocities}
\label{sect:velocities}
\begin{figure*}
\begin{center}
\includegraphics[width=0.45\hsize,angle=270]{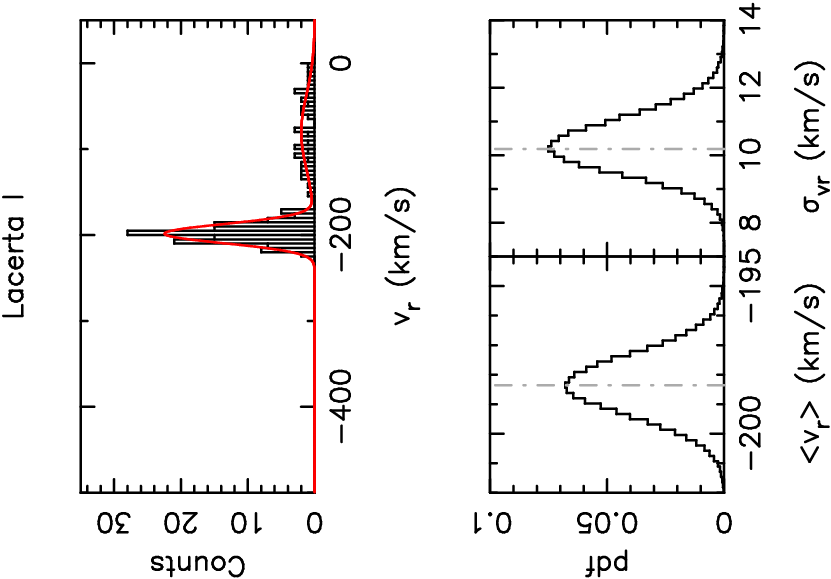}\hspace{0.5cm}
\includegraphics[width=0.45\hsize,angle=270]{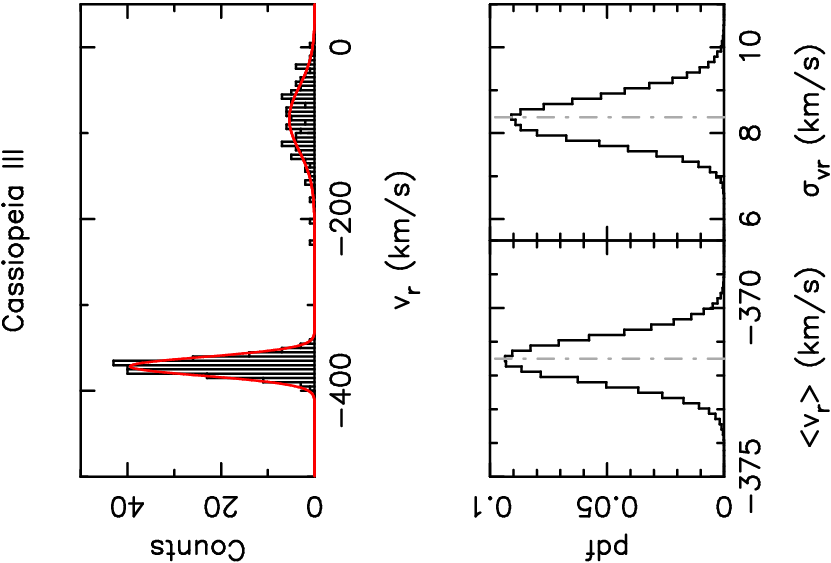}\hspace{0.5cm}
\includegraphics[width=0.45\hsize,angle=270]{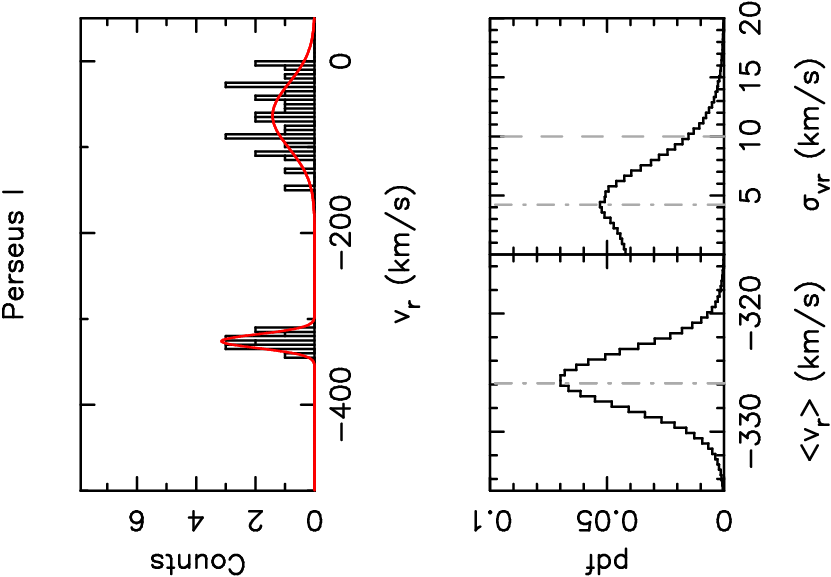}
\caption{\label{velocities}Kinematic information for all stars observed in Lac~I (left panels), Cas~III (middle panels), and Per~I (right panels). The top row shows the velocity histogram of the three velocity samples. In all cases, a clear clump of stars divulges the systemic velocity of the dwarf galaxy. The favored velocity models, as defined in equation (\ref{eq_model}) and convolved by the typical velocity uncertainty, are represented by the red lines. The bottom row of panels displays the marginalized pdf of the systemic velocity of a dwarf galaxy, $\langle v_\mathrm{r}\rangle$, and its velocity dispersion, $\sigma_\mathrm{vr}$. The dash-dotted line represents the favored model, i.e. the mode of the distributions, whereas the dashed line in the case of Per~I corresponds to the 90-percent confidence limit.}
\end{center}
\end{figure*}

The top row of panels in Figure~\ref{velocities} summarizes the kinematic information on the three dwarf galaxies. In all cases, the collapsed velocity histogram shows a clear detection of a dwarf galaxy's systemic velocity, well within expectations for M31 satellite dwarf galaxies ($-560<v_\mathrm{r}<-100\kms$; \citealt{tollerud12,collins13}) and well separated from the foreground Galactic contamination ($v_\mathrm{r}\simgt-150\kms$). In order to accurately derive the probability distribution functions (pdfs) of their systemic velocities and velocity dispersions, we determine the likelihood, given the data, of a family of models that is the sum of three Gaussian functions in velocity (one each for the Milky Way (MW) foreground contamination, the M31 halo contamination, and the dwarf galaxy). The likelihood of data point $i$, $d_i = \left\{v_{\mathrm{r},i},\delta_{\mathrm{vr},i}\right\}$, is defined as 

\begin{eqnarray}
\label{eq_model}
\ell_i\left(v_{\mathrm{r},i},\delta_{\mathrm{vr},i}|\mathcal{P}\right) = & \left(1-\eta_\mathrm{MW}-\eta_\mathrm{M31}\right)\mathcal{G}\left(v_{\mathrm{r},i}|\langle v_\mathrm{r}\rangle,\sqrt{\sigma_\mathrm{vr}^2+\delta_{\mathrm{vr},i}^2}\right)\nonumber\\
 & +\eta_\mathrm{MW}\mathcal{G}\left(v_{\mathrm{r},i}|\langle v_\mathrm{r,MW}\rangle,\sqrt{\sigma_\mathrm{vr,MW}^2+\delta_{\mathrm{vr},i}^2}\right)\nonumber\\
 & +\eta_\mathrm{M31}\mathcal{G}\left(v_{\mathrm{r},i}|\langle v_\mathrm{r,M31}\rangle,\sqrt{\sigma_\mathrm{vr,M31}^2+\delta_{\mathrm{vr},i}^2}\right).
\end{eqnarray}

\noindent with $\mathcal{G}(x|\mu,\sigma) = \frac{1}{\sqrt{2\pi}\sigma}\exp\left(-0.5\left((x-\mu)/\sigma\right)\right)$ and $\mathcal{P}=\left\{\langle v_\mathrm{r}\rangle,\sigma_\mathrm{vr},\eta_\mathrm{MW},\langle v_\mathrm{r,MW}\rangle,\sigma_\mathrm{vr,MW},\eta_\mathrm{M31},\langle v_\mathrm{r,M31}\rangle,\sigma_\mathrm{vr,M31}\right\}$ the eight parameters of the model. These correspond, in this order, to the systemic velocity of the dwarf galaxy, its velocity dispersion, the fraction of stars in the MW foreground model, the systemic velocity of this MW model and its velocity dispersion, the fraction of stars in the M31 contamination model, the systemic velocity of this M31 model and its velocity dispersion.

With the use of flat priors within physical ranges for all parameters\footnote{In particular, we enforce $-400<\langle v_\mathrm{r,M31}\rangle<-200\kms$, $\sigma_\mathrm{vr,M31}<150\kms$, and $\langle v_\mathrm{r,MW}\rangle>-180\kms$.}, we can directly calculate the probability of a model given the $N$ velocity measurements, $\overrightarrow{d}$, modulo a constant:

\begin{eqnarray}
P\left(\mathcal{P}|\overrightarrow{d}\right) \propto \prod\limits_{i=1}^N\ell_i\left(v_{\mathrm{r},i},\delta_{\mathrm{vr},i}|\mathcal{P}\right).
\end{eqnarray}

\noindent We use a Markov Chain Monte Carlo (MCMC) scheme following a Metropolis-Hastings algorithm to explore the parameter space. Finally, the integration of $P$ over the 6 MW and M31 nuisance parameters and either the dwarf galaxy's velocity dispersion or its systemic velocity yields the one-dimensional pdf of $\langle v_\mathrm{r}\rangle$ or $\sigma_\mathrm{vr}$, respectively. These are shown in the bottom row of panels in Figure~\ref{velocities} whereas the favored model, convolved with the typical velocity uncertainty ($5\kms$), is represented over the velocity histograms by the red line in the top row of panels.

\begin{table}
\caption{\label{properties}Spectroscopic properties of the three dwarf galaxies}
\begin{tabular}{l|ccc}
 & Lacerta~I & Cassiopeia~III & Perseus~I\\
\hline
$\alpha$ (ICRS)$^a$ & $22^\mathrm{h}58^\mathrm{m}16.3^\mathrm{s}$& $0^\mathrm{h}35^\mathrm{m}59.4^\mathrm{s}$ & $3^\mathrm{h}01^\mathrm{m}23.6^\mathrm{s}$\\
$\delta$ (ICRS)$^a$ & $+41\deg17'28''$ & $+51\deg33'35''$ & $+40\deg59'18''$\\
$L_V$ ($\lsun$)$^a$ & $10^{6.8\pm0.3}$ & $10^{6.6\pm0.3}$ & $10^{6.0\pm0.3}$\\
$\langle v_\mathrm{r}\rangle$ ($\kms$) & $-198.4\pm1.1$ & $-371.6\pm0.7$ & $-326\pm3$\\
$\langle v_\mathrm{r,gsr}\rangle$ ($\kms$) & $+9\pm2$ & $-186\pm2$ & $-220\pm4$\\
$\sigma_\mathrm{vr}$ ($\kms$) & $10.3\pm0.9$ & $8.4\pm0.6$ & $4.2^{+3.6}_{-4.2}$\\
$M_\mathrm{half}$ ($\msun$) & $4.2^{+0.8}_{-0.9}\times10^7$ & $4.1^{+0.7}_{-1.1}\times10^7$ & ---\\
$(M/L)_{V,\mathrm{half}}$ ($\msun/\lsun$) & $15^{+12}_{-9}$ & $8^{+9}_{-5}$ & ---\\
median $\FeH$ & $-2.0\pm0.1$ & $-1.7\pm0.1$ & ---\\
stacked $\FeH$ & --- & --- & $-2.0 \pm0.2$\\
\end{tabular}
$^a$Taken from \citet{martin13a} and \citet{martin13c}.
\end{table}

\begin{figure}
\begin{center}
\includegraphics[width=0.7\hsize, angle=270]{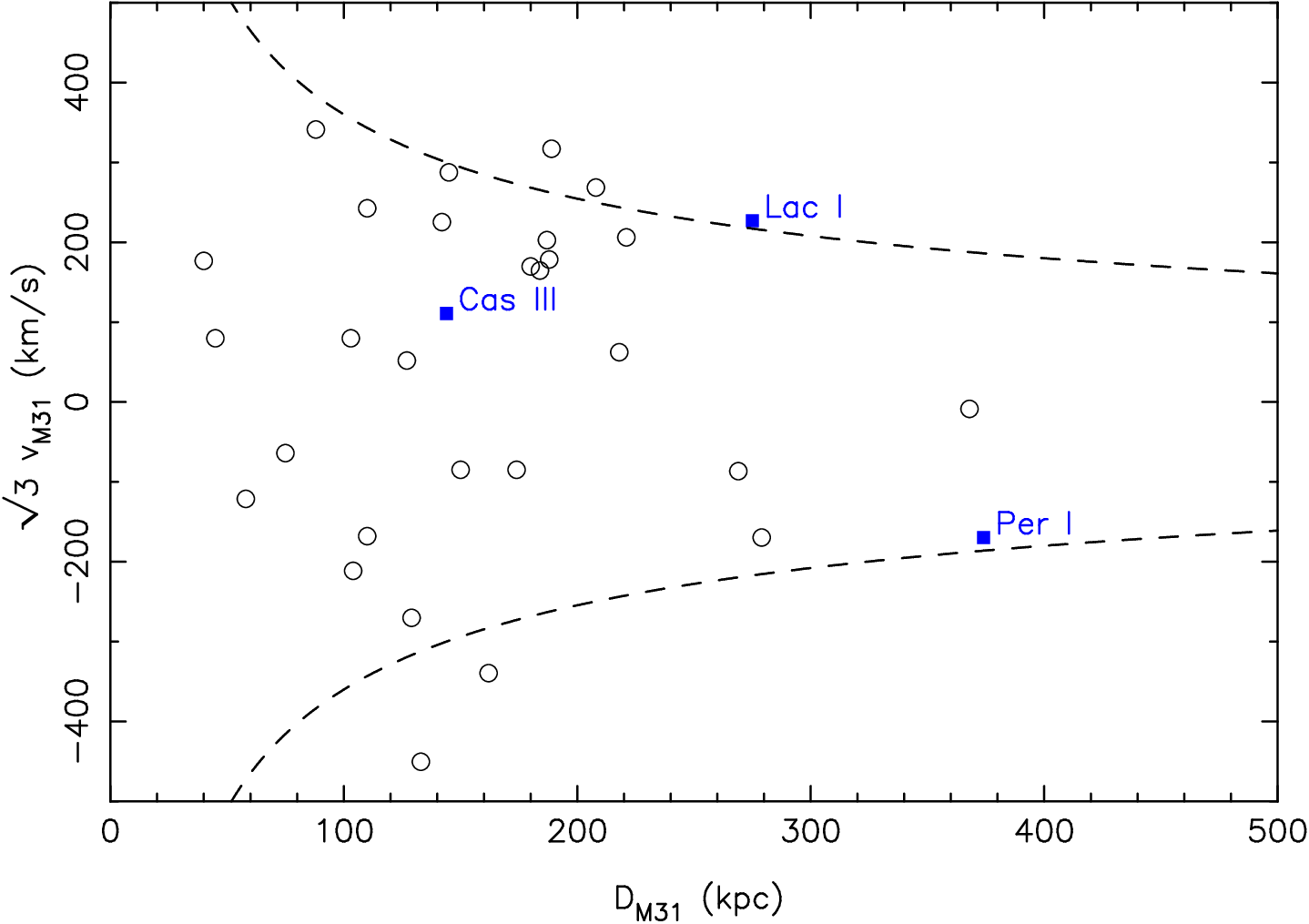}
\caption{\label{D_vrM31}Distribution of M31 satellite dwarf galaxies in the M31-centric velocity versus distance plane. Lac~I, Cas~III, and Per~I are highlighted in blue. Values for the other dwarf galaxies are taken from \citet{mcconnachie12}, \citet{conn12}, and \citet{collins13}. The dashed lines correspond to the escape velocity for a $1.5\times10^{12}\msun$ point-mass M31 \citep{watkins10}. Following \citet{mcconnachie12}, the $y$-axis values have been multiplied by $\sqrt{3}$ to account for the unknown transverse motion of the dwarf galaxies.}
\end{center}
\end{figure}

All three dwarf galaxies have well measured systemic velocities\footnote{The quoted uncertainties are determined from the parameter values at which $P\left(\mathcal{P}|\overrightarrow{d}\right)$ has dropped to 61 percent of the peak value, i.e. $\pm1\sigma$ deviations in the case of a Gaussian pdf.} with $\langle v_\mathrm{r}\rangle = -198.4\pm1.1\kms$ for Lac~I, $-371.6\pm0.7\kms$ for Cas~III, and $-326\pm3\kms$ for Per~I. These transform into the following Galactic standard of rest velocities \citep{dehnen98}: $\langle v_\mathrm{r,gsr}\rangle = +9\pm2\kms$, $-186\pm2\kms$, and $-220\pm4\kms$, respectively. Although the line of sight velocities at the large M31-distances of the three dwarf galaxies include part of the systemic proper motion of M31, they are still mainly dominated by the M31 systemic radial velocity. The difference, $v_\mathrm{r,M31}$, between $\langle v_\mathrm{r,gsr}\rangle$ and $\langle v_\mathrm{r,gsr}^\mathrm{M31}\rangle=-122\kms$ therefore informs us as to whether the dwarf galaxies are likely to be bound to M31. As can be seen in Figure~\ref{D_vrM31}, all dwarf galaxies are likely bound to M31 despite their large M31-centric distances, under the assumption that they have reasonable proper motions. Lac~I and Per~I could potentially be beyond the escape velocity of M31 if we are to assume that our cosmic neighbor has a low mass ($<1.5\times10^{12}\msun$), but this appears unrealistic given its large number of satellite dwarf galaxies, globular clusters, and stellar streams \citep[e.g.][]{martin13c,diaz14}.

The high quality of the Lac~I and Cas~III samples, with 126 and 212 member stars, respectively, also yields good constraints on their velocity dispersions, which we infer to be $\sigma_\mathrm{vr}=10.3\pm0.9\kms$ and $8.4\pm0.6\kms$. In the case of Per~I, the constraints on the velocity dispersion is much weaker since they rely on only 12 stars. The favored value is $\sigma_\mathrm{vr}=4.2\kms$, but only models with $\sigma_\mathrm{vr}>10\kms$ can reasonably be rejected (at the 90-percent confidence level). All the properties of the dwarf galaxies are summarized in Table~\ref{properties}.

Marginalizing over all models for a given dwarf galaxy, we also calculate the membership probability of every star in the sample for this dwarf galaxy. In all three cases, there is a clear separation between the contamination and the likely member stars, which all have very high membership probabilities.

We use equation (11) of \citet{walker09} to infer an estimate of the dynamical masses of Lac~I and Cas~III, folding in the uncertainties on the structural parameters (sampling from the MCMC chain of \citealt{martin13a}), those on the distance modulus \citep{martin13a}, and those on the velocity dispersion by sampling the MCMC chain calculated above. This yields $M_\mathrm{half} = 4.2^{+0.8}_{-0.9}\times 10^7\msun$ and $4.1^{+0.7}_{-1.1}\times 10^7\msun$ for Lac~I and Cas~III, respectively, and $V$-band dynamical-mass-to-light ratios within the half-light radius of $(M/L)_{V,\mathrm{half}} = 15^{+12}_{-9}$ and $8^{+9}_{-5}$ in solar units\footnote{The uncertainties on these $(M/L)_V$ values are mainly driven by the uncertainties on the luminosity of the systems.}.

\subsection{Metallicities}
In the case of Lac~I and Cas~III, we measure the metallicity, $\FeH$, of each member star using the equivalent widths of the Ca triplet lines. First, we determine the continuum for each star by fitting a B-spline to the wavelength region surrounding the triplet (8400--8700 \AA). We then divide the spectrum by the continuum fit, resulting in a normalised spectrum from which we can measure the equivalent widths of the Ca triplet by simultaneously fitting the continuum level using a polynomial and the 3 Ca lines using 3 Gaussian profiles. Each line fit is then checked for contamination from nearby sky lines by comparing the ratios of their equivalent widths with the expectation from higher resolution studies of the Ca triplet in RGB stars \citep[e.g.][]{battaglia08}. Contaminated lines are excluded from the determination of the final equivalent width for the triplet (see \citealt{collins13} for more details). Finally, we use the \citet{starkenburg10} relation to determine the value of $\FeH$ for each star from this equivalent width measurement\footnote{The reader is referred to \citet{collins13} for the details of the recalibration of this relation to use either 2 or 3 clean Calcium triplet lines.} and its $I$-band absolute magnitude, converted from the PS1 photometry \citep{tonry12}.

A comparison of the metallicity yielded by stars observed on multiple masks implies that the uncertainties on these individual $\FeH$ measurements are sizeable ($\sim~0.4$ dex), preventing a measure of the metallicity spread. We can nevertheless use the large size of the samples to derive a precise median metallicity. These are derived from a Monte Carlo sampling of the individual metallicity values and their associated uncertainties, yielding $\FeH = -2.0\pm0.1$ and $-1.7\pm0.1$ for Lac~I and Cas~III, respectively.

In the case of the lower quality Per~I data, we are limited to generating a single stacked spectrum of the 12 stars with a high membership probability. This spectrum yields the average metallicity $\FeH = -2.0\pm0.2$ for this system.

\section{Discussion}
\begin{figure}
\begin{center}
\includegraphics[width=0.7\hsize,angle=270]{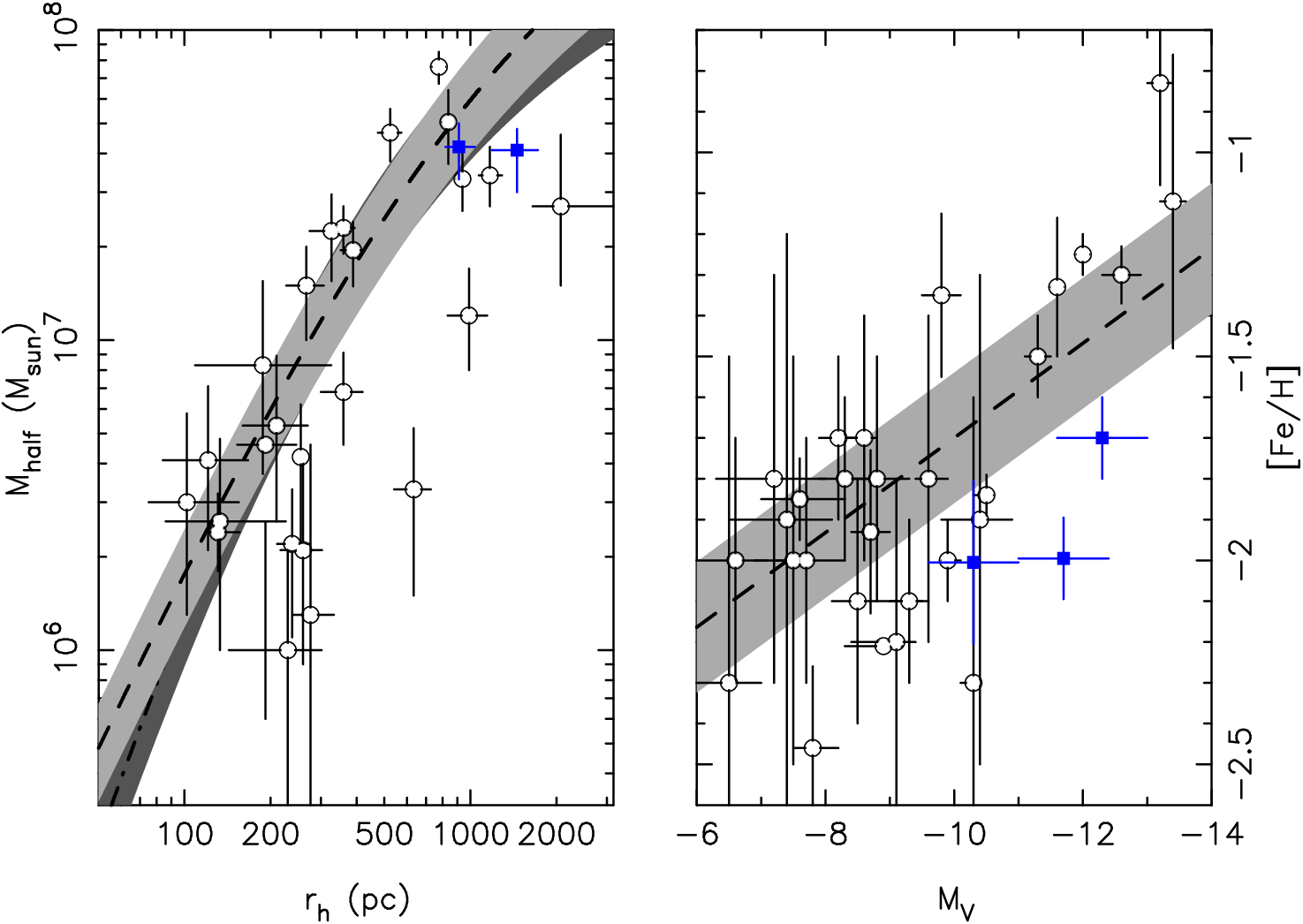}
\caption{\label{comparison}\emph{Left panel:} The mass-size relation of Lac~I and Cas~III (blue squares) compared to the other satellite dwarf galaxies of M31 (white dots). The values for the half-light radius are preferentially taken from the set of homogeneous PAndAS measurements (N. F. Martin et al., in preparation), completed by values taken from \citet{mcconnachie12}, \citet{bell11}, and \citet{slater11}. The mass calculations are taken from \citet{collins14}. The gray bands correspond to the universal mass profiles with scatter determined by \citet{collins14} for a cored profile (dark gray) and an NFW profile (light gray). \emph{Right panel:} The same for the luminosity-metallicity relation, with all three dwarf galaxies represented as blue squares. The metallicity values are spectroscopic metallicity measurements taken from \citet{collins13}, \citet{kirby13}, \citet{tollerud13}, and \citet{ho14}. The magnitudes are taken from the same references as the sizes.}
\end{center}
\end{figure}

We have presented the first spectroscopic study of the three dwarf galaxies Lac~I, Cas~III, and Per~I, recently found in the PS1 $3\pi$ survey. The systemic line-of-sight velocities of the three systems ($-198 < \langle v_\mathrm{r}\rangle < -372\kms$; $-220 < \langle v_\mathrm{r,gsr}\rangle < +9\kms$; $|v_\mathrm{r,M31}|<131 \kms$) confirm that they are indeed bound satellites of M31, except if they hold high tangential velocities (Figure~\ref{D_vrM31}). In the case of Lac~I and Cas~III, we obtain good constraints on their velocity dispersions, which we infer to be $\sigma_\mathrm{vr}=10.3\pm0.9\kms$ and $8.4\pm0.6\kms$. Combined with the known size of these two dwarf galaxies, these yield dynamical masses of $M_\mathrm{half}=4.2^{+0.8}_{-0.9}\times10^7\msun$ and $4.1^{+0.7}_{-1.1}\times10^7\msun$. In the case of Per~I, we infer a favored velocity dispersion of only $4.2\kms$, but the data is of too-low quality to confidently reject velocity dispersions below $10\kms$ at the 90-percent confidence level. Finally, we determine the average metallicity of the three dwarf galaxies; all three are metal-poor systems with $\FeH=-2.0\pm0.1$, $-1.7\pm0.1$, and $-2.0\pm0.2$, respectively.

All the properties we derive for the three dwarf galaxies are typical of other M31 (or Local Group) satellite galaxies, as we exemplify in Figure~\ref{comparison} in which we compare the dwarf galaxies with the other Andromeda dwarf galaxies in the context of the \citet[][left panel]{collins14} size-mass relation, and the luminosity-metallicity relation of \citet[][right panel]{kirby13}. The only point of mild tension comes from the metallicity of the  dwarf galaxies; these are on the low side of expectations. However, it should be pointed out that the luminosity-metallicity relation was determined from a combination of MW dwarf spheroidal galaxies and Local Group dwarf irregular galaxies (equation (3) of \citealt{kirby13}) and may not faithfully represent M31 satellites. It should also be noted that the metallicity of many of the M31 satellites are rather uncertain as they are usually measured from only a handful of stars. Overall, these do not significantly differ from our $\FeH$ measurements for Cas~III and Per~I. Lac~I appears more metal-poor than expected, but this could also be due to an overestimation of its (uncertain) luminosity. Confirmation of its departure from the luminosity-metallicity relation will have to be confirmed with deeper photometry than accessible by the PS1 survey.

While the left panel of Figure~\ref{comparison} focusses on the comparison of the size and mass of the dwarf galaxies with dark-matter dominated models, it is interesting to note that the velocity dispersion values we measure are in agreement with the MOND-ian predictions calculated by \citet{mcgaugh13}. The favored low velocity dispersion we infer for Per~I, albeit with large uncertainties, is also in agreement with similar predictions by \citet{pawlowski14}. 

Finally, it should be mentioned that, although Cas~III overlaps with the thin rotating disk of Andromeda dwarf galaxies \citep{ibata13a}, its velocity indicates that it is moving counter to the global motion of that structure.

A detailed study of the dynamics of Lac~I and Cas~III, based on their large velocity samples, will be the topic of a future contribution.

\acknowledgments

N.F.M. thanks Ana\"is Gonneau and Ariane Lan\c{c}on for their help in identifying carbon stars and gratefully acknowledges the CNRS for support through PICS project PICS06183.

The Pan-STARRS1 Surveys (PS1) have been made possible through contributions of the Institute for Astronomy, the University of Hawaii, the Pan-STARRS Project Office, the Max-Planck Society and its participating institutes, the Max Planck Institute for Astronomy, Heidelberg and the Max Planck Institute for Extraterrestrial Physics, Garching, the Johns Hopkins University, Durham University, the University of Edinburgh, Queen's University Belfast, the Harvard-Smithsonian Center for Astrophysics, the Las Cumbres Observatory Global Telescope Network Incorporated, the National Central University of Taiwan, the Space Telescope Science Institute, the National Aeronautics and Space Administration under Grant No. NNX08AR22G issued through the Planetary Science Division of the NASA Science Mission Directorate, the National Science Foundation under Grant No. AST-1238877, the University of Maryland, and Eotvos Lorand University (ELTE).

The data presented herein were obtained at the W.M. Keck Observatory, which is operated as a scientific partnership among the California Institute of Technology, the University of California and the National Aeronautics and Space Administration. The Observatory was made possible by the generous financial support of the W.M. Keck Foundation. The authors wish to recognize and acknowledge the very significant cultural role and reverence that the summit of Mauna Kea has always had within the indigenous Hawaiian community.  We are most fortunate to have the opportunity to conduct observations from this mountain

%


\end{document}